\documentclass[aps,prb,reprint,groupedaddress]{revtex4-1}
\usepackage[dvips]{graphicx}
\usepackage{amsmath}

\begin{document}

\title{Ultrafast and spatially resolved studies of charge carriers in atomically-thin molybdenum disulfide}

\author{Rui Wang}
\author{Brian A. Ruzicka}
\author{Nardeep Kumar}
\author{Matthew Z. Bellus}
\author{Hsin-Ying Chiu}\email{chiu@ku.edu}
\author{Hui Zhao}\email{huizhao@ku.edu}

\affiliation{Department of Physics and Astronomy, The University of Kansas, Lawrence, Kansas 66045, USA}

\date{\today}

\begin{abstract}
Atomically-thin molybdenum disulfide is emerging as a new nanomaterial with potential applications in the fields of electronic and photonics. Charge carrier dynamics plays an essential role in determining its electronic and optical properties. We report spatially and temporally resolved pump-probe studies of charge carriers in atomically-thin molybdenum disulfide samples fabricated by mechanical exfoliation. Carriers are injected by interband absorption of a 390-nm pump pulse and detected by measuring differential reflection of a time-delayed and spatially-scanned probe pulse that is tuned to an exciton transition. Several parameters on charge carrier dynamics are deduced, including carrier lifetime, diffusion coefficient, diffusion length, and mobility.
\end{abstract}

\maketitle

Molybdenum disulfide is a transition metal dichalcogenide with an indirect bandgap of 1.29~eV.\cite{GmelinHandbook} It is composed of monolayers of S-Mo-S that are bound by the weak van der Waals force, while the atoms in each layer are bound strongly by ionic-covalent interactions. The layered structure allows fabrication of atomically-thin films where the quantum confinement can significantly modify the electronic and optical properties. Although few-layer\cite{prsls27369,pr140A536,jap371928} and even monolayer\cite{mrl21457} MoS$_2$ samples have been fabricated and studied since 1960's, they have become topics of great interest again, partially owing to the progress in studies of graphene.\cite{pnas10210451,rpp74082501}

Very recently, strong photoluminescence (PL) was observed from few-layer MoS$_2$ samples,\cite{l105136805,nl101271} which was attributed to an indirect-to-direct band gap transition that occurs when varying the thickness from bulk to monolayer.\cite{l105136805,nl101271}  Such a transition was also confirmed by theoretical calculations\cite{b83245213} and scanning photoelectron microscopy measurements.\cite{b84045409}  In addition to the possible use in photonic applications, monolayer MoS$_2$ transistors with a 10$^8$ on/off ratio and a room-temperature mobility of more than 200~cm$^2$/Vs have been demonstrated.\cite{nn6147} This shows great potential in electronic applications,\cite{ieeeted583042,nl113768} since further improvements can be expected by, e.g. removing the ripples observed on monolayers.\cite{nl115148} Although samples used in these studies\cite{l105136805,nl101271,nn6147} were prepared by the simple mechanical exfoliation method,\cite{pnas10210451} techniques with potential for large-scale production, such as liquid and chemical exfoliations,\cite{s331568,am343944,cm233879} have been demonstrated to produce samples with high-mobility\cite{am234178} and similar PL properties.\cite{nl115111}

For its applications in photonics and electronics, it is necessary to understand the dynamics of charge carriers in atomically-thin MoS$_2$. Previous steady-state optical studies, including PL,\cite{l105136805,nl101271,nl115111,PSSR201105589} absorption,\cite{l105136805,nl115111} reflection,\cite{l105136805,nl101271} photoconductivity,\cite{l105136805} and Raman scattering,\cite{acsnano42659} have revealed many aspects of electronic and lattice properties. However, time-resolved optical measurements can provide direct information about the carrier dynamics, as illustrated by a recent time-resolved PL measurement.\cite{apl99102109}

Here we use a spatially and temporally resolved pump-probe technique to study charge carriers in few-layer MoS$_2$ samples fabricated by mechanical exfoliation. Carriers are injected by direct interband absorption of a 390-nm pump pulse, and detected by measuring the differential reflection of a 660-nm probe pulse. We found that the excitonic absorption is reduced by the carriers, which is consistent with the phase-state filling effect. By spatially and temporally resolving the signal, we deduce a carrier lifetime of 100$\pm$10 ps and a carrier diffusion coefficient of 20$\pm$10 cm$^2$/s, corresponding to a mobility of 800 cm$^2$/Vs and a diffusion length of 450 nm. The demonstrated technique can be used as a powerful tool to study charge carriers in various MoS$_2$-based structures.

\begin{figure}
 \includegraphics[width=8.5cm]{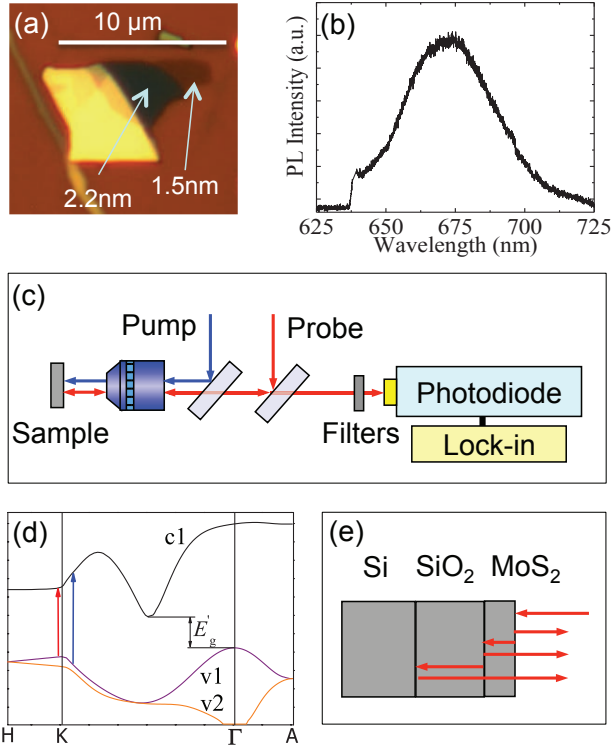}
 \caption{(a) Optical microscopy photo of the studied MoS$_2$ flakes under white light illumination. (b) PL spectrum of the sample. (c) Schematics of the optical pump-probe setup. (d) Schematics of the bandstructure of bulk MoS$_2$ and the pump/probe configuration. The blue and red vertical arrows indicate the pump and probe photon energies, respectively. (e) Geometry of the multiple reflection from the sample. The rays are vertically shifted for clarity. }
  \label{setup}
\end{figure}

Few-layer MoS$_2$ samples are fabricated by mechanical exfoliation\cite{pnas10210451} with an adhesive tape from natural crystals (SPI Supplies). By using a silicon substrate with a 90-nm SiO$_2$ layer, flakes with different atomic layers can be readily identified by using the contrast of microscopy images.\cite{Nanotechnology22125706,apl96213116} Figure \ref{setup}(a) shows an example of the identified flakes. From atomic-force microscopy measurements, one region has an average thickness of 1.5 nm. Since monolayer MoS$_2$ (one Mo layer sandwiched by two S layers) is 0.65-nm thick, we assign this region to a bilayer, considering uncertainties in the measurement. The region next to it (to the left) has an average thickness of 2.2 nm. It is assigned as a trilayer, which is next to a thick region of many layers (yellowish area). A PL spectrum obtained by focusing a 633-nm beam to the bilayer region is shown in Fig.~1(b). It peaks at about 675 nm. The sudden drop at the short-wavelength side is caused by a long-pass filter used to block the laser beam. Recent optical measurements have shown that the energy of the "A" exciton transition in atomically-thin samples is in the range of 655 - 685 nm.\cite{l105136805,nl101271,nl115111,apl99102109} Our result is consistent with these studies. When the excitation laser spot was moved to the trilayer region, no significant change in PL intensity was observed. This is a further confirmation that the first region is a bilayer instead of a monolayer, since PL from monolayer is expected to be much stronger than several-layer samples due to its direct bandgap.\cite{l105136805,nl101271} 

Figure \ref{setup}(c) shows schematically the optical pump-probe setup. The pump pulse is focused to the sample surface by using a microscope objective lens. The probe pulse with a different wavelength is focused by the sample lens, and its reflection is collected and detected by a photodiode, the output of which is detected by a lock-in amplifier. The pump and the probe pulses are linearly polarized along perpendicular directions. A mechanical chopper modulates the pump at about 2 kHz for lock-in detection. Reflection and scattering of the pump are prevented from reaching the photodiode by using color filters. In this configuration, the lock-in amplifier measures the change in the reflection of the probe pulse from the sample induced by the pump. Such a quantity is then normalized to provide a differential reflection, $\Delta R / R_0 = (R - R_0) / R_0$, where $R$ and $R_0$ are the reflection of the probe pulse from the sample with and without the presence of the pump pulse, respectively. All the measurements are carried out under ambient conditions and at room temperature.

Figure 1(d) shows schematically the bandstructure of bulk MoS$_2$.\cite{l105136805} Recent studies have shown that the indirect bandgap $E'_g$ increases from the bulk value of 1.29 eV to more than 1.90 eV in monolayer, while the direct excitonic transitions (from V1 and V2, respectively, to C1) at the {\bf K} point remain largely unchanged. We use a pump pulse of 390 nm (3.179 eV) to excite the electrons from the valance band to the conduction band [the blue vertical arrow]. The pulse is obtained by frequency doubling a 100-fs, 780-nm pulse from a Ti:sapphire laser with a repetition rate of 81 MHz. The 660-nm probe pulse is obtained by frequency doubling the 100-fs, 1320-nm output of an optical parametric oscillator that is pumped by the Ti:sapphire laser. The probe photon energy of 1.878 eV [the red vertical arrow] is near the "A" exciton transition from V1 to C1 at the {\bf K} point. According to the PL spectrum shown in Fig.~1(b), the 660-nm probe is within the "A" exciton transition but higher than the peak energy. It is lower than the "B" exciton transition from V2 to C1 of 2.04 eV (608 nm).\cite{l105136805} Hence, it detects the change of the "A" exciton transition induced by the carriers injected by the pump pulse.

\begin{figure}
 \includegraphics[width=8.5cm]{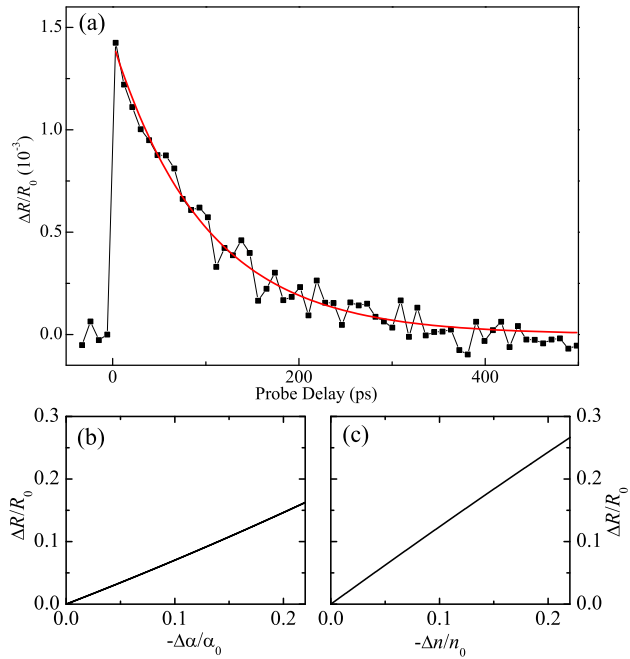}
 \caption{(a) Differential reflection signal measured with a 390-nm pump and a 660-nm probe. The solid line is a single exponential fit with a time constant of 100$\pm$10 ps. (b) Calculated $\Delta R/R_0$ as a function of $- \Delta \alpha / \alpha_0$ by using Eq.~1, over a rather large range of $\Delta R/R_0$. For small $\Delta R/R_0$ on the order of $10^{-3}$, the relation is approximately linear. (c) Calculated $\Delta R/R_0$ as a function of $- \Delta n / n_0$ by using Eq.~1. }
 \label{390pump}
\end{figure}

Figure 2(a) shows the measured differential reflection signal as a function of the probe delay, with a pump fluence of 11 $\mu$J/cm$^2$. We find that the signal reaches a peak instantaneously (pulse-width limited) and exists for several hundred ps. The signal is positive, and is on the order of 10$^{-3}$. The decay of the signal can be fit by a single exponential function [the red line in Fig.~\ref{390pump}(a)], with a decay time of 100$\pm$10 ps. 

In order to use the differential reflection to monitor the charge carriers, it is necessary to establish the relation between them. As shown in Fig.~1(e), the reflection beam is a superposition of three beams from the front surface and from the two interfaces. Hence, under normal incidence, the reflection coefficient of this multilayer system can be written as\cite{Nanotechnology22125706,apl96213116,b78235408}
\begin{widetext}
\begin{equation}
R(\tilde{n}_{1})=\left\vert \frac{r_{1}e^{i(\phi _{1}+\phi _{2})}+r_{2}e^{-i(\phi_{1}-\phi _{2})}+r_{3}e^{-i(\phi _{1}+\phi _{2})}+r_{1}r_{2}r_{3}e^{i(\phi_{1}-\phi _{2})}}{e^{i(\phi _{1}+\phi _{2})}+r_{1}r_{2}e^{-i(\phi _{1}-\phi_{2})}+r_{1}r_{3}e^{-i(\phi _{1}+\phi _{2})}+r_{2}r_{3}e^{i(\phi _{1}-\phi_{2})}}\right\vert ^{2},
\end{equation}
\end{widetext}
where $r_{1}=(\tilde{n}_{0}-\tilde{n}_{1})/(\tilde{n}_{0}+\tilde{n}_{1}), r_{2}=(\tilde{n}_{1}-\tilde{n}_{2})/(\tilde{n}_{1}+\tilde{n}_{2}), r_{3}=(\tilde{n}_{2}-\tilde{n}_{3})/(\tilde{n}_{2}+\tilde{n}_{3})$, and $\phi _{i}={2\pi d_{i}n_{i}}/{\lambda }$. In these formula, $\tilde{n}_1$, $\tilde{n}_2$, and $\tilde{n}_3$ are the complex indices of refraction for MoS$_2$, SiO$_2$, and Si, respectively. The $d_{i}$ are the thickness of the corresponding layers, and $\lambda$ is the wavelength. In general, the pump-injected carriers in MoS$_{2}$ will change both the real part and the imaginary part of the index of refraction, $\tilde{n}_{1} = n + \alpha i$, where $n$ and $\alpha$ are the real index of refraction and the absorption coefficient of MoS$_{2}$, respectively. These will change the reflection coefficient according to Eq. 1. 

Although the reflection coefficient depends on $n$ and $\alpha$ in a rather complex way, for small changes, the differential reflection is expected to be proportional to the changes in $n$ and $\alpha$. We use Eq. 1 to verify this. By using published values of the $n$ and $\alpha$ of thin-film MoS$_{2}$,\cite{ l105136805} we obtain $R_0$. We then vary $\alpha$ by a small quantity, $\Delta \alpha$, and use Eq.~1 to calculate the corresponding $\Delta R$. With this procedure, we obtain $\Delta R/R_0$  as a function of $\Delta \alpha / \alpha_0$. The result is plotted in Fig.~2(b). Although a slight deviation from a linear relation can be seen with relatively large $\Delta \alpha / \alpha_0$, for small variations,  $\Delta R / R_0$ is almost proportional to $\Delta \alpha / \alpha_0$. Since the measured $\Delta R/R_0$ is on the order of 10$^{-3}$, we can safely assume this linear relation. By using a similar procedure, we verify the almost-linear relation between $\Delta R/R_0$ and $\Delta n/n_0$, as shown in Fig.~2(c).  

In our experiments, the probe pulse is tuned to the strong excitonic absorption peaks. Hence, we expect the absorption change to be the major contribution to the signal. That is, although Figs. 2(b) and 2(c) show that the same relative changes in $n$ and $\alpha$ induce similar magnitudes of $\Delta R/R_0$, the pump-injected carriers cause a larger $\Delta \alpha / \alpha_0$ than $\Delta n / n_0$.  Because $\Delta R/R_0$ and $\Delta \alpha / \alpha_0$ have opposite signs, the positive signal shown in Fig.~2(a) indicates that the absorption of MoS$_2$ is decreased by the pump-injected carriers. This is consistent with the phase space filling effect of free carriers and excitons,\cite{b326601} which is often the dominant nonlinear effect observed in other materials like GaAs, although we could not exclude possible contributions from screening of the electron-hole interaction\cite{b326601} and bandgap renormalization.\cite{b322266} For saturated absorption, $\alpha = \alpha_0/(1-N/N_{sat})$, where $N$ and $N_{sat}$ are the carrier density and saturation density, respectively.\cite{bookboyd} Hence,  $\Delta \alpha / \alpha_0 = N / (N - N_{sat})$. Since the observed $\Delta \alpha / \alpha_0$ is on the order of $10^{-3}$, in these measurements $N \ll N_{sat}$ and $\Delta \alpha / \alpha_0 \propto N$, approximately.

With the established linear relation between the measured $\Delta R/R_0$ and the carrier density, we can directly monitor the carrier dynamics. We can now attribute the 100-ps decay time obtained from the fit shown in Fig.~\ref{390pump}(a) to the lifetime of the carriers excited by the pump pulse. We note that such a lifetime is consistent very well with the recently measured decay times of excitonic PL of about 100 ps.\cite{apl99102109} 

\begin{figure}
 \includegraphics[width=8.5cm]{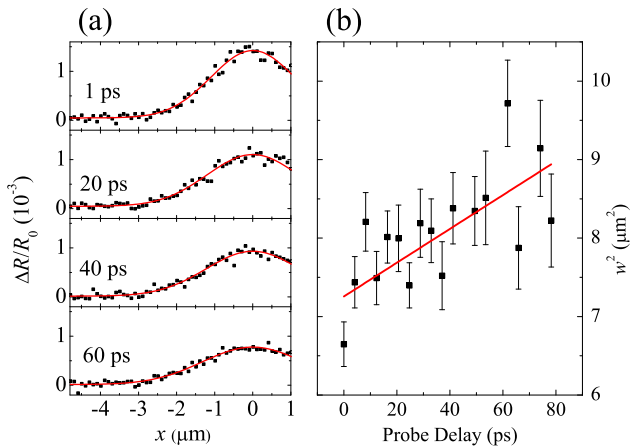}
 \caption{(a) Spatial profiles of the $\Delta R/R_0$ signal measured by scanning the probe spot with respect to the pump spot, with several probe delays as labeled in each panel. The solid lines are gaussian fits. (b) Square width of the spatial profiles deduced from gaussian fits as a function of the probe delay. The solid line is a linear fit, corresponding to a diffusion coefficient of 20$\pm$10 cm$^2$/s.}
 \label{diffusion}
\end{figure}

The relatively long lifetime makes it possible to study diffusion of the carriers out of the pump laser spot during their lifetime. For a classical diffusion process with a gaussian initial profile that is determined by the pump laser spot, the squared width (full width at half maxima) at time $t$ is  $w^2(t)=w^2(t_0)+16\mathrm{ln}2 D (t-t_0)$, where $w^2(t_0)$ is the squared width at an earlier time, and $D$ is the diffusion coefficient.\cite{b385788} Hence, by measuring the width of the carrier profile as a function of probe delay, we can deduce $D$. In this study, the produced flakes are rather small; hence, it is difficult to measure the full spatial profiles. Therefore, we position the laser spots close to the boundary of a flake, and acquire one side of the profile by scanning the probe spot with respect to the pump spot. A few examples with several probe delays are shown in the inset of Fig.~\ref{diffusion}(a). We fit the measured profiles by gaussian functions, and the deduced widths are plotted in Fig.~\ref{diffusion}(b). Despite the large uncertainties, a trend of increase of the width with time is observed. By a linear fit [the red line in Fig.~\ref{diffusion}(b)], we deduce a diffusion coefficient of 20$\pm$10 cm$^2$/s. With the carrier lifetime of $\tau$ = 100 ps, this gives a diffusion length of $L_D=\sqrt{D\tau}=450$~nm.

Owing to the Coulomb attraction, the excited electrons and holes move together in this diffusion process. Hence, the measured quantity is the ambipolar diffusion coefficient, $D_a = 2D_eD_h/(D_e+D_h)$, where $D_{e(h)}$ is the unipolar diffusion coefficient of electrons (holes).\cite{Neamenbook} Based on the fact that the effective masses of electrons and holes in MoS$_2$ are similar,\cite{ieeeted583042} we assume $D_e \approx D_h$. Hence, the measured value can be treated approximately as unipolar diffusion coefficients of electrons and holes. Using the Einstein relation, $D/k_BT = \mu/e$, where $k_B$, $T$, $\mu$, and $e$ are the Boltzmann constant, the temperature, the mobility, and the absolute value of the electron charge, the measured $D$ corresponds to a mobility of $\mu \approx 800$~cm$^2$/Vs. Here, we have assumed the carriers have a thermal distribution with a temperature of 300 K during the diffusion process. Since the measurement was performed over a probe delay range of about 100 ps, which is much longer than typical thermalization and energy relaxation time of a few ps in semiconductors, this assumption is valid.

It is interesting to compare the optically measured mobility with electric measurements. Initial electric measurements on exfoliated monolayer MoS$_2$ have indicated very low mobilities on the order of 1~cm$^2$/Vs.\cite{pnas10210451}. Very recently, it was found that by covering the MoS$_2$ layer with a dielectric layer with a high dielectric constant, such as HfO$_2$, the mobility can be increased to 780~cm$^2$/Vs for monolayer MoS$_2$.\cite{nn6147} This was attributed to suppression of Coulomb scattering by the dielectric\cite{nn6147,l98136805,nl92571} and modification of phonon dispersion.\cite{nn6147,nl62442} The MoS$_2$ flakes studied here are not covered by a dielectric layer. However, the mobility deduced is comparable to those with a dielectric top layer. We attribute this fact to the ambipolar nature of the diffusion process: In this process, the electron-hole pairs move as a unit. Since the pair is electrically neutral, it is less influenced by the Coulomb scattering. Therefore, although the MoS$_2$ sample is not covered by a dielectric layer, the mobility we measured indicates to the one that is not limited by the Coulomb scattering. We note that such a comparison is also limited by the exciton formation from the electron-hole pairs, which is enhanced in atomically-thin layers due to the quantum confinement. 

In summary, we have used a femtosecond pump-probe technique to study charge carriers in few-layer MoS$_2$ samples. Carriers are injected by direct interband absorption of a 390-nm pump pulse, and detected by measuring the differential reflection of a probe pulse of 660-nm. We found that the absorption is reduced by the carriers, which is consistent with the phase-state filling effect. By spatially and temporally resolving the differential reflection signal, we deduce a carrier lifetime of 100$\pm$10 ps and a carrier diffusion coefficient of 20$\pm$10 cm$^2$/s, corresponding to a mobility of 800 cm$^2$/Vs and a diffusion length of 450 nm. Besides providing quantitative information on the carrier dynamics in this new promising nanomaterial, our experiment may stimulate further optical studies of carrier dynamics in this material system. The optical approach has the advantage of being noninvasive, avoiding device fabrication and influence of electric contacts.

We would like to thank Chih-Wei Lai and Shenqiang Ren for many helpful discussions, and Rodolfo Torres-Gavosto and Cindy Berry for their help on AFM measurements. We acknowledge support from the US National Science Foundation under Awards No. DMR-0954486 and No. EPS-0903806, and matching support from the State of Kansas through Kansas Technology Enterprise Corporation. Acknowledgment is also made to the Donors of the American Chemical Society Petroleum Research Fund for support of this research.


%

\end{document}